\documentclass[12pt]{JHEP}

\usepackage{epsfig}
\epsfclipon
\usepackage{multicol}
\usepackage{epsfig,bm}
\usepackage{amssymb,amsmath}

\def\be{\begin{equation}}
\def\ee{\end{equation}}
\def\bea{\begin{eqnarray}}
\def\eea{\end{eqnarray}}

\def\pref#1{(\ref{#1})}

\def\beq{\begin{equation}}
\def\eeq{\end{equation}}
\def\beqa{\begin{eqnarray}}
\def\eeqa{\end{eqnarray}}

\def\cA{{\cal A}}

\def\cT{{\cal T}}

\def\cG{{\cal G}}

\newcommand{\bmat}{\left(\begin{array}}
\newcommand{\emat}{\end{array}\right)}

\def\yzero{\smash{\hbox{$y\kern-4pt\raise1pt\hbox{${}^\circ$}$}}}

\def\-{\hphantom{-}}

\def\s2{\frac{1}{2}}

\def\IF{\relax{\rm I\kern-.18em F}}
\def\II{\relax{\rm I\kern-.18em I}}
\def\IP{\relax{\rm I\kern-.18em P}}
\def\IC{\relax{\rm I\kern-.48em C}}
\def\IR{\relax{\rm I\kern-.18em R}}
\def\IK{\relax{\rm I\kern-.20em K}}
\def\IM{\relax{\rm I\kern-.25em M}}

\def\cA{{\cal A}}

\def\Dsl{\,\raise.15ex\hbox{/}\mkern-13.5mu D} %this one can be subscripted
\def \one{\relax{\rm 1\kern-.26em I}}

\newcommand{\ba}{\begin{eqnarray}}
\newcommand{\ea}{\end{eqnarray}}

\newcommand{\comment}[1]{}

\newcommand{\hs}{\hspace{-1.12mm}}
\newcommand{\eq}{\hs & = & \hs}

\newcommand{\boldk}{{\boldsymbol{k}}}
\newcommand{\boldp}{{\boldsymbol{p}}}

\newcommand{\boldr}{{\boldsymbol{r}}}

\newcommand{\boldy}{{\boldsymbol{y}}}
\newcommand{\bolds}{{\boldsymbol{s}}}

\newcommand{\bzero}{{\boldsymbol{0}}}

\def\dphi{\hat{\phi}}
\def\dPhi{\hat{\Phi}}

\def\dh{\hat{h}}

\title{Deviations From Newton's Law in\\Supersymmetric Large Extra Dimensions}

\author{P.~Callin$^1$ and C.P.~Burgess$^{2,3}$
\\

$^1$ Institute of Physics, University of Oslo, N-0316 Oslo, Norway.\\
${}^2$ Department of Physics and Astronomy, McMaster University,\\
\qquad 1280 Main Street West, Hamilton, Ontario, Canada, L8S 4M1.\\
${}^3$ Perimeter Institute, 31 Caroline Street North,\\ \qquad
Waterloo, Ontario, Canada.\\
E-mail: \email{n.p.callin@fys.uio.no},
\email{cburgess@perimeterinstitute.ca}\\

}

\date{}
\abstract{Deviations from Newton's Inverse-Squared Law at the
micron length scale are smoking-gun signals for models containing
Supersymmetric Large Extra Dimensions (SLEDs), which have been
proposed as approaches for resolving the Cosmological Constant
Problem. Just like their non-supersymmetric counterparts, SLED
models predict gravity to deviate from the inverse-square law
because of the advent of new dimensions at sub-millimeter scales.
However SLED models differ from their non-supersymmetric
counterparts in three important ways: ($i$) the size of the extra
dimensions is fixed by the observed value of the Dark Energy
density, making it impossible to shorten the range over which new
deviations from Newton's law must be seen; ($ii$) supersymmetry
predicts there to be more fields in the extra dimensions than just
gravity, implying different types of couplings to matter and the
possibility of repulsive as well as attractive interactions; and
($iii$) the same mechanism which is purported to keep the
cosmological constant naturally small also keeps the
extra-dimensional moduli effectively massless, leading to
deviations from General Relativity in the far infrared of the
scalar-tensor form. We here explore the deviations from Newton's
Law which are predicted over micron distances, and show the ways
in which they differ and resemble those in the non-supersymmetric
case.}

\begin{document}

%\newpage

%===================================================================================

\section{Introduction}

The recent discovery of a non-vanishing Dark Energy density
\cite{DEdiscovery} throws into sharp relief the general
theoretical befuddlement about how to predict reliably the
gravitational response to the energy of the vacuum
\cite{ccreview}. In a nutshell, the difficulty arises because in
four dimensions gravity responds to the energy density of the
vacuum, $\rho$, as if it were a cosmological constant. But $\rho$
generically receives its largest contributions from the quantum
zero-point energy of the shortest-wavelength modes, for instance
with modes having wavelength $\ell$ contributing in four
dimensions an amount of order $\delta \rho \sim \ell^{-4}$. As
such, the theoretical prediction is sensitive to the most
microscopic of a theory's details, and since any theory describing
the known particle types has $\ell^{-1}$ larger than 100 GeV, the
theoretical prediction is generically many orders of magnitude
larger than the observed value:\footnote{From here on we adopt
units for which $\hbar=c=1$.}
\be
    \rho_{\rm obs} = \frac{\hbar c}{a^{4}} \qquad \hbox{with}
    \qquad \frac{\hbar c}{a} \sim 3 \times 10^{-3} \quad \hbox{eV} \,.
\ee

\subsection{Naturalness Issues}

Until the observational result was found to be nonzero, the theoretical goal
had been to identify a theory which is both experimentally successful and {\it
technically natural} in that sense that the contributions to the vacuum energy
for some reason cancel once summed over the relevant particle content. Now that
we know the result is nonzero the bar has been raised, and a technically
natural solution would instead require a cancellation only for those modes
having scales $\ell < a$. Although this kind of naturalness is not absolutely
required by fundamental principles, it amounts to the requirement that the
small hierarchy can be understood within an effective theory of microscopic
physics, regardless of the scales for which this theory is formulated.
Technical naturalness is a conservative requirement because we know that it
applies for {\it all} of the many other hierarchies of scale for which we have
solid evidence, in all branches of physics.

Yet it is hard to modify physics at scales above $\ell \sim a$ to
get a small enough vacuum energy without also running into
conflict with the many experimental measurements which are
available for such scales. More recently, despair at making
progress along these lines has led many to abandon the criterion
of technical naturalness altogether and to turn to anthropic ideas
along the lines proposed by Weinberg more than ten years ago
\cite{anthropic,anthropic2}. This point of view has gained
additional momentum from the observation that anthropic reasoning
might be required to make predictions within string theory, given
the enormous number of vacua which arise there \cite{landscape}.
This line of argument has emboldened others to abandon using
technical naturalness as a criterion for understanding the
electro-weak hierarchy \cite{splitsusy}.

On the other hand, potential progress has also been made in obtaining a small
gravitational response to the vacuum energy without abandoning technical
naturalness, within the framework of Supersymmetric Large Extra Dimensions
(SLED) \cite{Towards,Doug,SLEDreviews}. The idea behind this proposal is that
it is indeed possible to modify physics at scales $\ell \lesssim a$ in such a
way as only to modify gravitational interactions without changing
non-gravitational physics. As a result, although the vacuum energy may be large
it may be possible to modify how this vacuum energy gravitates for scales $\ell
< a$ without also ruining the agreement between theory and the vast number of
non-gravitational experiments. The framework within which this can be done is
that of large extra dimensions \cite{ADD}, according to which two internal
dimensions are taken to be as large as they can possibly be without coming into
conflict with observations. Interestingly enough, the largest radius which is
possible for such extra dimensions turns out to be $r \sim a$, as imposed by
tests of Newton's Law at short distances \cite{NLTests,NLTestsRev}.

It is proposed that within such models supersymmetry in the extra dimensions
can play the crucial role of cancelling the contributions to the vacuum energy
from all modes having wavelengths $\ell \lesssim r \sim a$. There is currently
considerable activity devoted to understanding what the gravitational response
to the vacuum energy would look like within this framework, in order to see
whether this kind of modification really provides a technically natural
description of the dark energy
\cite{Towards,Doug,SLEDreviews,Precursors,SLEDrelated}.

\subsection{The Smoking Gun}

A particularly attractive feature of the SLED framework is that it is very
predictive, and so eminently falsifiable. This predictiveness ultimately comes
because technical naturalness requires the proposal to modify physics at length
scales $\ell \lesssim a$ (or energies above $a^{-1} \sim 10^{-3}$ eV) --- a
range of scales to which we have a great deal of experimental access. In
particular SLED models add to the observational implications for the Large
Hadron Collider (LHC) of the earlier large-extra-dimensional models
\cite{LEDPheno}, as well as predicting new features to do with supersymmetry
\cite{preSLEDPheno,MSLED,SLEDPheno}. As such, SLED models provide a unique
framework which links the technical naturalness of the vacuum energy with
observational implications at high-energy accelerators.

Given that its {\it raison d'etre} is modifying the gravitational response of
the vacuum, it is the modifications to gravity which provide the most
definitive signatures for the SLED proposal. These come in two types,
corresponding to deviations from General Relativity over both sub-millimeter
and cosmological distance scales. Over very large distances the model behaves
like a scalar-tensor theory, with the light scalar (or scalars) describing the
dynamics of the moduli of the extra dimensions. Unlike for non-supersymmetric
models having large extra dimensions, within SLED models there is an
understanding of why these modes remain essentially massless in a technically
natural way, which is a consequence of the same mechanism which is purported to
protect the vacuum energy in these models \cite{ABRS}. This leads to
implications for the time-dependence of the Dark Energy density
\cite{ABRS,SLEDCosmo}, as well as model-dependent implications for
long-distance tests of gravity.

However, it is tests of Newton's Law over micron length scales
which ultimately provide the smoking gun for SLED models. This is
because SLED models rely for their success on the existence of two
large extra dimensions, and on the coincidence between the size,
$r$, of the extra dimensions and the observed scale, $a$, in the
observed Dark Energy density. This means that Newton's law {\it
must} be violated at length scales, $\lambda$, which are of order
the Kaluza Klein (KK) masses for the large extra dimensions:
$\lambda \sim r/2\pi \sim a/2\pi$, which works out to be around a
micron in size \cite{MSLED}. Since the size of these dimensions
cannot be shrunk without making the observed vacuum energy too
large, there is no way to escape the implication that Newton's law
must change once tested over these distances.

Since these deviations are predicted over distances which are not
too far from experimental reach, we can hope that the theory can
be definitively tested within the not-to-distant future. For these
purposes it is important to have precise predictions for what
kinds of deviations should be expected. It is the purpose of this
paper to provide a first calculation of the short-distance
behaviour of force laws within the SLED picture, working within
the framework of the simplest (toroidal) compactifications. In
particular, we compute the forces which should be expected between
bodies which are separated in the visible dimensions but are not
displaced in the extra dimensions, due to the mediation of the
various bosons which propagate within the extra dimensions. As
might be expected, there are more of these bosons than arise in
non-supersymmetric models due to the additional particle content
which supersymmetry in the bulk requires. Some of these additional
bosons can mediate spin-dependent forces, and the detection of
such could provide a way to distinguish between the supersymmetric
from non-supersymmetric options.

We organize our results as follows. The next section, \S2, briefly
describes the bosonic field content of 6D supergravity theories,
and then describes the quantization of their linearized
fluctuations about flat space. The exchange of these fluctuations
is then used in \S3 to compute the interaction between
slowly-moving, localized classical sources, and identify their
interaction potential energy. We also identify what kinds of
charges such sources can carry in the static limit, and follow how
the interaction energy depends on these charges. Finally, we close
in \S4 with a concluding discussion.

\section{Forces Mediated by 6D Bosons}

In this section we describe the bosonic fields which make up the
bulk sector in models with supersymmetric large extra dimensions.
We then identify how these bulk particles can couple to brane
degrees of freedom, and compute the forces which are obtained when
they are exchanged by two sources which are both localized on the
same brane.

\subsection{6D Supergravity}

The field content of practical interest for SLED models is the
bosonic sector of 6D supergravity coupled to various forms of
super-matter. In (2,0) supersymmetry the bosonic field content of
the supergravity multiplet consists of the metric, $g_{MN}$, and a
2-form gauge potential which is subject to a six-dimensional
self-duality condition. This is normally combined with a tensor
multiplet, whose bosonic content contains another 2-form potential
having the opposite self-duality property, and so which combines
with the previous one into an unconstrained potential, $B_{MN}$.
The other boson in the tensor multiplet is a scalar field, $\phi$,
known as the dilaton. The most common types of (2,0) matter
multiplets are gauge multiplets --- whose bosonic parts consist of
the gauge potentials, $A^\alpha_M$, of some gauge group --- or
hypermultiplets --- whose bosonic part consists of scalar fields,
$\Phi^a$. These latter fields parameterize a coset space, $G/H$,
and so nominally describe the Goldstone modes for the symmetry
breaking pattern $G \to H$. For some supergravities a scalar
potential, $v(\Phi)$, can exist for these fields, in which case
the symmetry $G$ is explicitly broken and the fields $\Phi^a$
become instead pseudo-Goldstone particles. Alternatively, if a
subgroup of $G$ is gauged, then some of these hyper-scalars can be
eaten by the gauge fields through the usual Anderson-Higgs-Kibble
mechanism.

For a broad class of supergravities these bosons are governed by
the following action\footnote{Our metric is `mostly plus' and we
adopt Weinberg's curvature conventions \cite{WbgG&C}, which differ
from those of Misner, Thorne and Wheeler \cite{MTW} only in an
overall change of sign of the Riemann tensor.}
\bea \label{BosonicAction}
  S &=& \int d^6 x \sqrt{-g} \left[
    -\frac{1}{2 \, \kappa^2} \, g^{MN} \Bigl( R_{MN} +
    \partial_M \phi \, \partial_N \phi + G_{ab}(\Phi) \, \partial_M
    \Phi^a \, \partial_N \Phi^b \Bigr) \right. \nonumber\\
    && \left. \qquad\qquad\qquad - \frac{1}{12} \, e^{-2{\phi}} \, G_{MNP} G^{MNP}
    - \frac{1}{4} \, e^{-{\phi}} \, F^\alpha_{MN} F_\alpha^{MN}
%    - \frac{2 g^2}{\kappa^4} \, e^{\phi} \, v(\Phi)
  \right] ,
  \label{eq:SalamSezgin_general}
\eea
where $\kappa = M^{-2}$ with $M$ being the 6D Planck mass, which
is of order 10 TeV \cite{NS}.
%$g$ is a coupling constant having dimensions
%of inverse mass, $g = \hat{g} \, M^{-1}$. For ungauged
%supergravities $g=0$ and so there is no scalar potential. For $g
%\ne 0$ the potential is nonzero, and the quantity $v(\Phi)$ can
%have either sign for different kinds of supergravity. We here
%envision $v \ge 0$, such as is true for chiral versions of 6D
%supergravity.
The quantity $G_{ab}(\Phi)$ represents a $G$-invariant metric on
the coset space $G/H$, and we here focus our attention on the case
where no scalar potential exists. Supersymmetry puts a number of
restrictions on the form the target-space metric, $G_{ab}$ can
take, but these do not play a role in what follows. Finally,
$F^\alpha_{MN} = \partial_M A^\alpha_N - \partial_N A^\alpha_M +
{c^\alpha}_{\beta\gamma} \, A^\beta_M \, A^\gamma_N$ is the usual
gauge-covariant field strength, and the field strength, $G_{MNP}$,
for the 2-form potential, $B_{MN}$, is given by
\be
  G_{MNP} = \partial_M B_{NP} +
    \partial_N B_{PM} + \partial_P B_{MN} +
    \tilde{c}\,\kappa \, \Omega_{MNP} \,,
\ee
where $\Omega_{MNP}$ is the Chern-Simons form, given for abelian
gauge groups by $\Omega_{MNP} = F_{MN} A_{P} + F_{NP} A_{M} +
F_{PM} A_N$. $\tilde{c}$ here is a supergravity-specific constant,
which can vanish for some of the extra-dimensional tensors
appearing in the bulk supergravity.

The equations of motion obtained from the action,
\pref{BosonicAction}, are given by Einstein's equation,
\bea \label{eq:SSeom_g}
  \frac{1}{\kappa^2} \, R_{MN}
  &=& \partial_M \phi \, \partial_N \phi + G_{ab}(\Phi) \,
  \partial_M \Phi^a \, \partial_N \Phi^b +
    \frac{1}{2} \, e^{-2{\phi}} \, G_{MAB} {G_N}^{AB} -
    e^{-{\phi}} \, F_{MA} {F_N}^B \nonumber \\
  && \qquad - \, \frac{1}{4} \, g_{MN} \left[
    \frac{1}{3} \, e^{-2{\phi}} \, G_{ABC} G^{ABC}
    - \frac{1}{2} \, e^{-{\phi}} \, F_{AB} F^{AB}
%    + \frac{4 g^2}{\kappa^4} \,  e^{{\phi}} \, v(\Phi)
  \right] \,,
\eea
supplemented by the other field equations
\bea
  0 &=& \frac{1}{\kappa^2} \, \Box \phi
    + \frac{1}{6} \, e^{-2{\phi}} \, G_{MNP} \, G^{MNP}
    + \frac{1}{4} \, e^{-{\phi}} \, F_{MN} F^{MN}
%    - \frac{2 g^2}{\kappa^4} \, e^{{\phi}} \, v(\Phi)
    \label{eq:SSeom_phi} \\
    0 &=& \Box \Phi^a + \Gamma^a_{bc}(\Phi)\,  g^{MN}
    \, \partial_M \Phi^b \, \partial_N \Phi^c
%    + \frac{2 g^2}{\kappa^4} \, e^\phi \, G^{ab}(\Phi) \,
%    \partial_b v
    \label{eq:SHyperscalar} \\
  0 &=&  \nabla_P
    \left( e^{-2{\phi}} \, G^{PMN} \right)  \label{eq:SSeom_B} \\
  0 &=& \nabla_N \left( e^{-{\phi}} \, F^{NM} \right) +
   \kappa \, e^{-2{\phi}} \, G^{MAB} F_{AB}  \,.
    \label{eq:SSeom_A}
\ea
Here the $\Gamma^a_{bc}(\Phi)$ represent the Christoffel symbols
built in the usual way from the target-space metric,
$G_{ab}(\Phi)$.

As noted earlier, these expressions do not include a potential for the scalar
fields, as would be appropriate for ungauged 6D supergravity. Even for ungauged
supergravity it could also happen that a scalar potential is generated for
fields like $\phi$ if some of the gauge field strengths, $F_{MN}$ or $G_{MNP}$,
are nonzero in the background. We do not consider here potentials coming from
either sources since the success of the SLED mechanism relies on these
potentials not being generated with a size which is large enough to be relevant
to our present focus: the forces between macroscopic objects over
sub-millimeter distances.

\subsection{Toroidal Compactification}

The simplest solution to these equations is the trivial one, for
which the scalars are constants while all gauge field-strengths
vanish: $F_{MN}^\alpha = 0$ and $G_{MNP} = 0$. Since these
assumptions ensure there is no matter stress-energy, the simplest
metric configuration is flat space: $g_{MN} = \eta_{MN}$. We
compactify to 4 dimensions by taking two of these flat directions
($x^m, m=4,5$) to fill out a compact 2-torus, $\cT_2$, of volume
$\cA$, while the other four dimensions ($x^\mu, \mu = 0,...,3$)
are large.

For these purposes we define $\cT_2$ by identifying points on the
flat plane according to
\begin{equation}\label{rr1}
    (x^4,x^5) \cong (x^4 + n_2 r_2 \cos\theta + n_1 r_1;
    x^5 + n_2 r_2 \sin\theta) \,,
\end{equation}
where $n_{1,2}$ are integers and $\theta$, $r_1$ and $r_2$ are the
three real moduli of the 2-torus. Equivalently, in terms of the
complex coordinate $z = x^4 + i x^5$ this is
\begin{equation}\label{rr}
    z \cong z +(n_2 \tau + n_1) r_1 \,,
\end{equation}
with the complex quantity $\tau$ defined by $\tau = (r_2/r_1) \,
e^{i\theta} = \tau_1+i \tau_2$, and as defined $\tau$ takes values
in the upper half-plane, $\tau_2 > 0$. In terms of these
quantities, the volume of the 2-torus becomes $\cA = r_1 r_2
\sin\theta$.

To set notation, we now dimensionally reduce a free 6D scalar
field on such a background. Consider therefore a 6D scalar field
$\psi$ whose 6D field equation is\footnote{The subscript `$6$'
here emphasizes that the d'Alembertian appearing here is the
6-dimensional one.} $(-\Box_6 + m^2) \, \psi$, propagating on a
spacetime compactified to 4 dimensions on the above 2-torus. We
assume the scalar satisfies the following boundary conditions
\begin{eqnarray} \label{bcnd}
    \psi(x^\mu,x^4 + n_2 r_2 \cos\theta + n_1 r_1;
    x^5 + n_2 r_2 \sin\theta)=
    e^{2\pi i(n_1 \,\rho_{1} +n_2 \,\rho_{2})}\,
    \psi(x^\mu,x^4,x^5) \,,
\end{eqnarray}
with $0 \le \rho_{1,2} < 1$ being two real quantities. The choices
$\rho_{1,2} = 0,\frac12$ correspond to taking periodic or
anti-periodic boundary conditions along the torus' two cycles,
although more general values of $\rho_i$ are also possible.

The scalar field $\phi$ may be expanded in terms of the
eigenfunctions of the 2D Laplacian, $\Box_2 = \partial_4^2 +
\partial_5^2$, corresponding to
\be \label{KKScalarExpansion}
   \psi(x, \boldy) =
    \sum_\boldp \psi_\boldp(x) \, u_\boldp(\boldy) \,,
\ee
where we denote the 4D coordinates, $x^\mu$, collectively by $x$ and the 2D
coordinates, $x^m$, as $\boldy$ (with $y^1 = x^4$ and $y^2 = x^5$). The mode
functions are given by plane waves, $u_\boldp(\boldy) = \cA^{-1/2} \, e^{i
\boldp \cdot \boldy}$, where the above boundary conditions imply the allowed
extra-dimensional momenta are given by
\begin{equation}\label{wavef}
    \boldp \cdot \boldy = 2 \pi \left[
    \frac{(n_1+\rho_1) (y^1 - y^2 \cot\theta)}{r_1}
    + \frac{(n_2+\rho_2) y^2 \csc \theta}{r_2} \right]\,.
\end{equation}
These modes satisfy $\Box_2 u_\boldp = - \boldp^2 \, u_\boldp$
with
\begin{eqnarray}\label{mass22}
    \boldp^2 &=&
    \frac{(2 \pi)^2}{\cA\, \tau_2} \, \Bigl| n_2+\rho_{2}-\tau
    (n_1+\rho_{1})\Bigr|^2 \,.
\end{eqnarray}

Viewed from a 4D perspective, each of the KK modes, $\psi_\boldp$, is a 4D
scalar field satisfying the field equation
\be
    (- \Box_4 + \boldp^2 + m^2 ) \, \psi_\boldp = 0 \,,
\ee
and so whose 4D mass is related to the 6D mass, $m$, of the full
6D field $\psi$, by $M^2_\boldp = \boldp^2 + m^2$. Similar
decompositions also hold for free higher-spin fields compactified
on a 2-torus, as is described in more detail (as needed) below.

\subsection{Linear Fluctuations}

The above free-field expressions are useful for the present
purposes because they may be applied to the dynamics of small
fluctuations about the flat vacuum solution described above. The
free-field results describe the leading (linear) parts of these
fluctuations. In particular, we wish to explore the potential
energy which is set up by various sources through the exchange of
small fluctuations in the bosonic fields which arise in 6D
supergravity, described earlier.

To this end we expand the action, \pref{BosonicAction}, about the
toroidal background configuration described above, working to
quadratic order in the deviations of the fields $\phi$, $\Phi^a$,
$A^\alpha_M$, $B_{MN}$ and $g_{MN}$ about this background. For an
appropriate choice for the various gauge-averaging terms (more
about which below) this leads to a quadratic action within which
fields of differing spins do not mix:
\be
    S_{\rm quad} = S_\phi + S_\Phi + S_A + S_B + S_h \,.
\ee
Each term in this action is described in more detail in the
following sections.

\subsubsection*{The Dilaton}

In the absence of background gauge field strengths, $F_{MN} =
G_{MNP} = 0$, there is no potential for the dilaton and so any
constant value $\phi_0$ provides an equally good background value.
Dividing the dilaton into background and fluctuation according to
\beq
    \phi = \phi_0 + \kappa \, \dphi \,,
\eeq
and expanding the action to quadratic order in the fluctuation,
$\dphi$, then leads to
\be
    S_\phi = -\frac12 \int d^6x \; \eta^{MN}
    \Bigl( \partial_M \dphi \, \partial_N \dphi \Bigr)
    \,.
\ee

Since this has the same form as the example discussed above of
dimensional reduction of a free scalar field on $\cT_2$ (in the
special case $m=0$), the results of this earlier section may be
taken over in whole cloth. In particular we decompose $\dphi$ in
terms of toroidal modes as in eq.~\pref{KKScalarExpansion},
leading to the tower of KK masses given in eq.~\pref{mass22}. The
momentum-space propagator for such a 6D field is then
\beq
    D^\phi(k^\mu,\boldp) = \frac{1}{k^M k_M}
    = \frac{1}{k^2 + \boldp^2} \,,
\eeq
where $k^M = \{k^\mu,\boldp\}$, with $k^\mu$ denoting the
(time-like) 4-momentum in the four visible directions (with 4D
contraction $k^2 = k^\mu k_\mu$) and $\boldp$ representing the
(Euclidean) 2-momentum in the compact two dimensions (with 2D
contraction $\boldp^2 = \boldp \cdot \boldp$).

The dilaton can couple directly to brane matter, depending on the
nature of the microscopic physics of the brane. For instance for
$D$-branes in string theory, branes typically couple (in the
absence of background gauge and Kalb-Ramond fields) to the dilaton
with strength
\beq \label{DilatonCoupling}
    S_b = \int d^d\xi \; \sqrt{- \gamma} \; e^{\lambda \phi} \, T\,,
\eeq
where $T$ is the brane tension, $d$ is the spacetime dimension of
the brane, $\lambda$ is a constant whose value depends on the kind
of brane involved, and $\xi^\mu$ are coordinates on the brane
world-sheet which we take to sweep out the surface $x^M(\xi)$ in
spacetime. Also, $\gamma = \det \gamma_{\mu\nu}$ where
$\gamma_{\mu\nu} = g_{MN} \partial_\mu x^M \partial_\nu x^N$
denotes the induced metric on the brane world-sheet. For some
branes it can happen that $\lambda$ vanishes, as it does for
instance in the case of $D$3-branes in the 10D Einstein frame.

To linear order in the fluctuation $\dphi$ we parameterize the
dilaton interaction with the form
\beq \label{LinearDilatonCoupling}
    S_{\rm int} =  \kappa \int d^d\xi \;
    \sqrt{-\gamma} \; \dphi \, \rho \,,
\eeq
where $\rho$ denotes the local source density. For instance for a
point particle of mass $m$ on a 3-brane one might have $\rho =
\lambda [T +  m \delta^3(\xi)]$, for some mass $m$. In general,
such a coupling to the brane tension, $T$, can ruin the existence
of the flat background solution of interest here due to the
back-reaction which the presence of a configuration of branes
induces on the intervening 6D bulk fields. Indeed, an
understanding of how the bulk adjusts is central to the issue of
whether 6D supergravity provides a naturally small dark-energy
density. In what follows we imagine that this back-reaction can be
negligible to the linearized order to which we work, and focus on
the implications for experiments of the dilaton couplings to
localized particles.

\subsubsection*{Goldstone Modes}

The discussion of the hyper-scalars, $\Phi^a$, follows a similar
route, once the fields are expanded into background plus
fluctuation according to
\beq
    \Phi^a = \varphi^a + \kappa \dPhi^a \,.
\eeq
The main difference is in this case the nature of the couplings to
brane-bound particles which might be expected.

In general, for constant background fields, $\partial_M \varphi^a
= 0$, it is always possible\footnote{We assume here the target
space manifold to have Euclidean signature.} to rescale the
fluctuations $\dPhi^a$ such that $G_{ab}(\varphi) = \delta_{ab}$,
in which case the quadratic action for $\dPhi^a$ becomes
equivalent to $N$ copies of the dilaton action, $S_\phi$:
\be
    S_\Phi = -\frac12 \int d^6x \; \delta_{ab} \, \eta^{MN}
    \partial_M \dPhi^a \, \partial_N \dPhi^b
    \,.
\ee
This then leads to the following momentum-space propagator for
$\dPhi^a$
\beq
    D^\Phi_{ab}(k^\mu,\boldp)
    = \frac{\delta_{ab}}{k^2 + \boldp^2} \,,
\eeq
where $k^2$ and $\boldp^2$ are defined as before.

Because $\Phi^a$ arises as a Goldstone boson its couplings are
typically derivative couplings, which precludes there being a
dilaton-like coupling of the form \pref{DilatonCoupling} directly
to brane matter. Instead, we expect the linearized coupling to
have the general form
\beq \label{LinearizedGBCoupling}
    S_{\rm int} = \kappa \int d^d\xi \;
    \sqrt{-\gamma} \; j^M_a \partial_M \dPhi^a \,,
\eeq
where $j^M_a$ denotes the contribution of background brane fields to the
conserved current of the symmetry for which $\Phi^a$ is the Goldstone boson.

\subsection{Gauge Fields}

To quadratic order the expansion of the gauge-field part of the
supergravity action about $A^\alpha_{M} = 0$ is equivalent to that
for a collection of non-interacting (abelian) gauge potentials,
\be
    S_A = - \int d^6x \; \left[ \frac14 F^\alpha_{MN}
    F_\alpha^{MN} + \frac{1}{2 \xi_1}
    (\partial^M A^\alpha_M )^2 \right]
    \,,
\ee
where a convenient covariant gauge-fixing Lagrangian has been added to remove
the gauge freedom $\delta A_M^\alpha = \partial_M \epsilon^\alpha$. The gauge
and Lorentz indices are raised and lowered using the metrics
$\delta_{\alpha\beta}$ and $\eta_{MN}$, respectively.

This then leads to the following momentum-space propagator for
$A_M^\alpha$
\beq
    D^{\alpha\beta}_{MN}(k^\mu,\boldp)
    = \frac{\delta^{\alpha\beta}}{k^2 + \boldp^2} \left[ \eta_{MN} +
    (\xi_1 - 1) \, \frac{k_M k_N}{k^2 + \boldp^2} \right] \,.
\eeq

The coupling of any such a 6D gauge boson to a brane again depends
on the microscopic details. If so, then we expect the linearized
coupling to have the general form
\beq \label{LinearizedGaugeCoupling}
    S_{\rm int} = \int d^d\xi \;
    \sqrt{-\gamma} \; J^M_\alpha A_M^\alpha \,,
\eeq
where we take the current, $J^M_\alpha$, to include the
appropriate 6D gauge coupling, $g$.

One should also be alive to more complicated possibilities,
however, such as if an abelian bulk gauge field, $F^\alpha_{MN}$,
were to mix with an abelian brane gauge field,
$V^\alpha_{\mu\nu}$, as in
\beq
    S_{\rm int}' = k_{\alpha\beta} \int d^d\xi \; \sqrt{-\gamma} \;
    F^\alpha_{\mu\nu} V^{\beta\mu\nu} \,,
\eeq
where $k_{\alpha\beta}$ represent a matrix of mixing parameters.
Phenomenological constraints can be quite strong for such
interactions, because if they involve the electromagnetic field
they can produce effects which are easily detected even if very
small \cite{FVTerms}.

\subsection{2-Form Gauge Potentials}

A field which is ubiquitous to higher-dimensional supergravity theories is the
skew-symmetric gauge potential, $B_{MN}$, since this field is usually related
to the graviton by supersymmetry. The quadratic action for this field in the
absence of background field strengths is
\bea
    S_B &=& - \int d^6x \; \left[ \frac{1}{12} \,G_{MNP}
    G^{MNP} + \frac{1}{2 \, \xi_a}
    (\partial_M B^{MN} )^2 \right] \nonumber \\
    &=& -  \int d^6x \; \left[\frac14 \partial_M B_{NP}
    \partial^M B^{NP} + \frac12\left( \frac{1}{\xi_a} - 1\right)
    \left( \partial_M B^{MN} \right)^2 \right]
    \,,
\eea
where $G_{MNP} = \partial_M B_{NP} + \partial_N B_{PM} +
\partial_P B_{MN}$ and the second term is the gauge-fixing
Lagrangian which is added to remove the gauge freedom $\delta B_{MN} =
\partial_M \Lambda_N - \partial_N \Lambda_M$, for arbitrary $\Lambda_M$. The
corresponding momentum-space propagator for $B_{MN}$ becomes
\beqa
    D^B_{MN,PQ}(k^\mu,\boldp)
    &=& \frac{1}{k^2 + \boldp^2} \left[
    \left( \eta_{MP} \eta_{NQ}
    - \eta_{MQ} \eta_{NP} \right)  \phantom{\frac12}\right.  \\
    && \; \left.+
    \left( \frac{\xi_a - 1}{k^2 + \boldp^2} \right)
     \, \left( \eta_{MP}k_Nk_Q - \eta_{NP}k_Mk_Q + \eta_{NQ} k_Mk_P -
     \eta_{MQ}k_Nk_P \right) \right] \,. \nonumber
\eeqa

The field $B_{MN}$ can couple to branes, although these couplings
often vanish in the absence of background field strengths in the
bulk or on the brane. An example is the $B_{MN}$ coupling to
$D$-branes, which is obtained (in the absence of Ramond-Ramond
fields) by substituting $G_{MN} \to g_{MN} + \kappa B_{MN} +
\kappa \alpha' V_{MN}$ inside $\gamma_{\mu\nu}$ in
eq.~\pref{DilatonCoupling}, where $\alpha'$ is a dimensionful
constant and as before $V_{MN} = \partial_M V_N - \partial_N V_M$
is the field strength of a brane gauge field.

Other types of antisymmetric tensors can also arise, which are not
related to the extra-dimensional metric by supersymmetry. Examples
of these are the Ramond-Ramond fields which arise within Type IIB
supergravity models in 10 dimensions. (Higher-rank fields can also
arise, although we do not pursue the implications of these further
here.)

In general, for slowly varying fields we expect a linearized
coupling to have the general form
\beq \label{LinearizedKRCoupling}
    S_{\rm int} = \kappa \int d^d\xi \;
    \sqrt{-\gamma} \; J^{MN} B_{MN} \,,
\eeq
for some model-dependent current, $J^{MN}= - J^{NM}$.

\subsection{Gravity}

Finally, we follow standard practice and expand the metric about
the background using
\be
  g_{MN} = \eta_{MN} + 2 \kappa \,  \dh_{MN} \,.
\ee
Expanding the Einstein-Hilbert action to quadratic order in
$\dh_{MN}$ then leads to
\bea
  S &=& - \int d^6 x \; \left[
    - \, \frac{1}{2} \partial_M \dh \, \partial^M \dh
    + \frac{1}{2} \partial_N \dh_{PQ} \, \partial^N \dh^{PQ}
    + \partial_P {\dh^{PN}} \, \partial_N \dh
    - \partial_P {\dh^{PN}} \, \partial_Q {\dh^Q}_{N} \right.\nonumber \\
    && \qquad\qquad\qquad \left.
    - \frac{1}{\xi_2} \, \left(\partial^M \dh_{MN} - \frac12
    \partial_N \dh \right)^2
  \right] \,, \nonumber\\
  \label{eq:pert_gzero}
\eea
where $\dh = \eta^{MN} \dh_{MN}$. With this choice the propagator
in momentum space (in 6 dimensions) is
\beqa
    D^h_{MN,PQ}(k^\mu,\boldp) \eq \frac{1}{k^2+\boldp^2} \left[
    \frac{1}{2} \left(
      \eta_{MP} \eta_{NQ} +
      \eta_{MQ} \eta_{NP}
    \right) - \frac{1}{4} \eta_{MN} \eta_{PQ}
  \right. \\
  && \hspace{4mm} \left. +
    \frac{(\xi_2-1)}{2 (k^2 + \boldp^2)} \left(
      \eta_{MP} k_N k_Q +
      \eta_{MQ} k_N k_P +
      \eta_{NP} k_M k_Q +
      \eta_{NQ} k_M k_P
    \right)
  \right] \,. \nonumber
\eeqa

The interaction between these fields and matter is the usual
gravitation one, with $\dh_{MN}$ coupling to the linearized stress
tensor:
\be \label{LinearizedGravityCoupling}
  S_\mathrm{int} = -\kappa \int d^d x \;
    T^{MN} \dh_{MN} \,.
\ee

\section{Interactions Amongst Non-relativistic Sources}

In this section we compute the interaction energy which is
generated by the linearized tree-level exchange of 6D bulk fields
between two classical charge distributions. For practical
applications we focus on the case where the sources sit on the
same brane, and so are separated only in the 3 visible spatial
directions and not separated at all in the 2 compact dimensions.

Consider now the exchange of bulk fields from classical sources
described by the couplings of eq.~\pref{LinearDilatonCoupling},
\pref{LinearizedGBCoupling}, \pref{LinearizedGaugeCoupling},
\pref{LinearizedKRCoupling} and \pref{LinearizedGravityCoupling}.
Such an exchange leads to the momentum-space interaction potential
\ba \label{Vofkdef}
    V(k) &=& \kappa^2 \Bigl[
    \rho_1(k) D^\phi(k) \rho_2(-k) + k^M {j_1}_M^a(k)
    D^{\Phi}_{ab}(k) k^N {j_2}_N^b(-k) \nonumber \\
    && \qquad +
    {J_1}^M_\alpha(k) D^{\alpha\beta}_{MN}(k) {J_2}^N_\beta(-k)
    %\eqnl
    + \, J_1^{MN}(k) D^B_{MN,PQ}(k)
    J_2^{PQ}(-k) \nonumber \\
    && \qquad \qquad +
    T_1^{MN}(k) D^h_{MN,PQ}(k) T_2^{PQ}(-k) \Bigr]\,,
  \label{eq:interaction_pot}
\ea
where $k^M = \{k^\mu,\boldp \}$ represents the 6-momentum transfer
which occurs due to the bulk-field exchange. Although all indices
written here are 6-dimensional in practice the macroscopic sources
on the brane have currents which are parallel to the branes, which
we take also to be the large 4 dimensions $x^\mu$.

For precision tests of gravity our interest is in the elastic scattering of
macroscopic non-relativistic sources, and so the momentum transfer which is
relevant in the large four dimensions is $k^\mu = (0, \boldk)$. We also keep
the extra-dimensional momentum transfer, $\boldp$, arbitrary because here we
know the brane states are localized at the brane positions, and so the
brane-bulk couplings in themselves break momentum conservation in these
directions.

\subsection{Point Sources}

In order to apply the above expression to real measurements of
forces we must ask what the source currents, $\rho$, $j^a_M$,
$J^\alpha_M$, $J_{MN}$ and $T_{MN}$, would be for a brane-bound
source. Although we might expect the answer in general to be
model-dependent, a considerable amount can be learned from
symmetries in the case when the interaction is non-relativistic
and elastic and the source is labelled simply by its 4-momentum
and spin. (For instance, in what follows we apply these results to
the interactions of electrons or protons or any other
non-relativistic object whose internal structure is not disturbed
by the bulk-field exchange.) In this case currents like $J_\mu =
\langle p,s|\hat J_\mu |p,s\rangle$ depend only on the particle
4-momentum $p^\mu$ and the spin pseudo-vector $s^\mu$, where
$p^\mu p_\mu = - m^2$, $p^\mu s_\mu = 0$ and $s^\mu s_\mu = + 1$,
and so parity and Lorentz invariance determines their form up to
an overall scalar normalization.\footnote{We impose parity
invariance on the matrix elements since the sources of interest
are bound states of parity-conserving interactions like
electromagnetism, the strong force or gravity.}

For instance, in such a case the single-particle matrix element of
a 4-dimensional vector, axial vector, skew-tensor, skew axial
tensor and symmetric tensor current must have the form
\beqa \label{FormFactors}
    &&J_\mu = q \, p_\mu \,, \qquad J^A_\mu = q_A \, s_\mu \,,\nonumber\\
    &&J_{\mu\nu} = c \, \epsilon_{\mu\nu\lambda\rho} p^\lambda
    s^\rho \,, \qquad J^A_{\mu\nu} = c_A (p_\mu s_\nu - p_\nu s_\mu)
    \nonumber\\
    \hbox{and} \quad &&T_{\mu\nu} = A p_\mu
    p_\nu + B s_\mu s_\nu \,,
\eeqa
where $q$, $q_A$, $c$, $c_A$, $A$ and $B$ are unknown constants
whose form is not determined by 4D Lorentz invariance.

Furthermore, for static interaction energies our interest is in negligible
velocities, so we may specialize to the rest frame for which $p^\mu \simeq (m,
\bzero)$ and $s^\mu = (0, \bolds)$, where $\bolds$ is the unit vector pointing
in the direction in which the particle's spin angular momentum is measured. In
this case the $m$ dependence of the various constants may be determined on
dimensional grounds, leaving only a dimensionless number undetermined.
Additionally, since all the currents $J^A_\mu$, $J_{\mu\nu}$ and $J^A_{\mu\nu}$
give the same kind of contributions to the interaction potential, we only keep
$J^A_{\mu\nu}$ for simplicity.

Using these arguments we find that the quantities which appear in
eq.~\pref{Vofkdef} are approximately constant for the range of momentum
transfers of interest, $\lambda(k) \approx \lambda(0)$ etc., and thus
\ba
  \lambda(0) &=& a  m \, , \nonumber \\
  j^\mu_a(0) &=& Q_a \, p^\mu/m \simeq
    Q_a  \, \delta^\mu_{\;0} \, , \nonumber \\
  J^\mu_\alpha(0)  &=&  g_\alpha q_\alpha p^\mu/m \simeq
    g_\alpha q_\alpha \, \delta^\mu_{\;0} \, , \nonumber \\
  J^{\mu\nu}(0)  &=&  \frac{c}{2}
    \left( p^\mu s^\nu - p^\nu s^\mu \right)
    = \frac{c}{2} \, m s^i \, \left( \delta^\mu_0 \, \delta^\nu_i
    - \delta^\nu_0 \delta^\mu_i \right) \,, \nonumber \\
  T^{\mu\nu}(0) &=& m \Bigl( \delta^\mu_{\;0} \delta^\nu_{\;0}
  + b \, s^i \, s^j \, \delta^\mu_i \delta^\nu_j \Bigr) \,.
\ea
Now $a$, $Q_a$, $q_\alpha$, $b$ and $c$ are {\it dimensionless} constants,
$g_\alpha$ is the 6D gauge coupling constant (having dimensions of inverse
mass) and we have used the definition of mass to fix the normalization of $m$
in the coefficient $A$ defined in eq.~\pref{FormFactors}.

Inserting these expressions into the interaction potential of
eq.~\pref{Vofkdef} (and specializing for simplicity to a single
gauge charge, $q$) then gives
%
%\be
%  V(k^\mu,\boldp) = \left\{ \kappa^2 {m_1 m_2} \left[
%    a_1 a_2
%    - c_1 c_2 (\bolds_1 \cdot \bolds_2) + \frac{3}{4}
%  \right] - g^2 {q_1} {q_2} \right\} \frac{1}{k^2 + \boldp^2} \,,
%\ee
\be
  V(k^\mu,\boldp) = \left\{ \kappa^2 {m_1 m_2} \left(
    X + \frac{3}{4}
  \right) - g^2 {q_1} {q_2} \right\} \frac{1}{k^2 + \boldp^2} \,,
\ee
where
\be
  X = a_1 a_2 - c_1 c_2 (\bolds_1 \cdot \bolds_2) +
    \frac{1}{4} \left( b_1 + b_2 \right) + b_1 b_2 \left[
      (\bolds_1 \cdot \bolds_2)^2 - \frac{1}{4}
    \right] ,
\ee
and $k^\mu = (k^0, \boldk)$. In this expression the first term ($a_1 a_2$)
arises due to dilaton exchange, the last term ($q_1 q_2$) from the exchange of
bulk gauge fields, the spin-dependent term ($c_1 c_2$) from the exchange of the
Kalb-Ramond 2-form potential, and the ($b_{1,2}$) terms pluss the penultimate
term ($3/4$) follows from 6D graviton exchange. Derivative couplings preclude
Goldstone boson exchange from contributing in the static limit.

Taking the limit $k^0 \to 0$ corresponding to elastic (static)
interactions and transforming to position space we then find
\ba
    V(r,\boldy) &=& \frac{1}{\cA} \int \frac{d^3 k}{(2\pi)^3}
    \sum_\boldp
    V(\boldk, \boldp) \, e^{i \boldk\cdot\boldr + i\boldp \cdot\boldy}
      \nonumber \\
    &=& - \frac{1}{\cA} \left\{ \kappa^2 m_1 m_2 \left(
    X + \frac{3}{4}
    \right) - g^2 {q_1} {q_2} \right\}
    \sum_\boldp \int \frac{d^3 k}{(2\pi)^3}
    \frac{e^{i\boldk\cdot\boldr + i\boldp\cdot\boldy}}{\boldk^2 + \boldp^2}
    \nonumber \\
    &=& -\frac{1}{4\pi r\cA} \left\{ \kappa^2 m_1 m_2 \left(
    X + \frac{3}{4}
    \right) - g^2 {q_1} {q_2} \right\}
    \sum_\boldp e^{-|\boldp| r + i\boldp\cdot\boldy} \,,
    \nonumber\\
\ea
where $r = |\boldr|$. Taking two sources which share the same brane we take
$\boldy = \bold0$, and see that the resulting $V(r)$ becomes the expected sum
over Yukawa potentials, one for each KK mode.

For sufficiently large $r$ only the zero mode $\boldp = \bold0$ contributes,
giving the standard $1/r$ result
\be \label{masslesspotential}
    V(r) \to  -\frac{\kappa^2 m_1 m_2}{4\pi r \cA}
    \left(X + \frac{3}{4} \right)
    + \frac{g^2 {q_1} {q_2}}{4 \pi r \cA} \,,
\ee
and from this we read off the four-dimensional Newton constant, $G_N$, by
focussing on the $b$-independent gravitational term (last term in the
parenthesis):
\be \label{eq:NewtonsConstant}
    G_N = \frac{3 \kappa^2}{16\pi \cA} \,.
\ee
Similarly, the effective gauge coupling constant, $e$, is given by
\beq
    \alpha = \frac{e^2}{4 \pi} = \frac{g^2}{4 \pi \cA} \,.
\eeq

Notice that although eq.~\pref{eq:NewtonsConstant} pre-multiplies
just the gravitational contribution (so defined because it
includes only the contribution of the massless modes of the
extra-dimensional metric) it differs from what would have been
obtained --- namely $G_N = \kappa^2/(8\pi \cA)$ --- using only the
4D graviton. This difference arises because of the contribution in
eq.~\pref{masslesspotential} of other 6D metric polarizations
besides the 4D graviton.

We see here another reflection of the fact that {\it
long-distance} gravity also deviates from General Relativity in
these extra-dimensional models, because of the existence within
them of massless KK metric modes which can mediate long-range
interactions. Although these same modes also exist in
non-supersymmetric models with large extra dimensions, it is
important to notice that in the non-supersymmetric case quantum
corrections lift their masses to the generic KK mass scale,
thereby removing any such long-distance signature. The same need
{\it not} be true within SLED, since in this case the suppression
of the quantum contributions to the vacuum energy also suppress
the corrections to the light-scalar masses, leaving these modes
naturally light enough to have observable effects over long
distances \cite{ABRS}. In this sense long-distance modifications
of General Relativity potentially distinguish supersymmetric from
non-supersymmetric large-extra-dimensional models, although the
strength of the couplings of these modes depends on the details of
the extra dimensions and how they have evolved over cosmological
timescales.

Although we find in the present instance more than one such
massless KK mode, this is an artefact of our having compactified
from 6D to 4D on a torus. Generic compactifications need have only
two light scalar fields, corresponding to the dilaton arising from
the approximate scale invariance of 6D supergravities together
with its axionic partner under 4D supersymmetry. It is these
fields which are ultimately responsible for the time-dependence of
the Dark Energy density which SLED models predict
\cite{ABRS,SLEDCosmo}.

Another torus-specific feature of the above results is the fact
that the KK spectrum is identical for bulk fields having different
spins. For general extra-dimensional geometries these spectra can
differ, leading to a more complicated interplay of the strengths
of different fields as a function of source separation.

\begin{figure}[h!]
  \centering
  \includegraphics{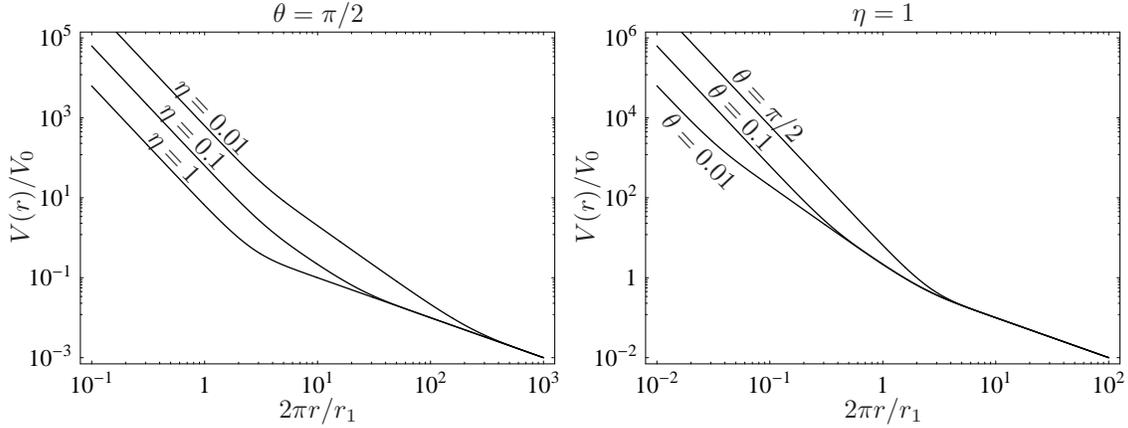}
  \caption{
  The potential in eq.~\pref{eq:potential_torus_6d} calculated numerically
  for different choices of the moduli $\theta$ and $\eta = r_1/r_2$
  (where $r_2$ is defined to be the larger of the two toroidal radii), with $
  V_0 / (2\pi) = -({G_N m_1 m_2}/{r_1}) \left( 1 + \frac{4X}{3} \right) +
  ({ \alpha {q_1} {q_2} }/{r_1})$. The large- and small-$r$ regimes where
  $V(r) \sim 1/r$ and $1/r^3$ respectively is clearly seen, as is an
  intermediate regime for which $V(r) \sim 1/r^2$ for very asymmetric torii
  ($\eta=0.01$ (left figure) and $\theta=0.01$ (right figure)).
  }
  \label{fig:numeric_results}
\end{figure}

With these definitions we finally obtain the following interaction
potential energy for two point sources situated at the same point
in the extra dimensions but separated by a distance $r$ in the
large 3 space dimensions.
\beq
    V(r) = \left\{ -\frac{G_N m_1 m_2}{r} \left(
    1 + \frac{4X}{3} \right) + \frac{ \alpha {q_1} {q_2} }{r}
  \right\} \sum_\boldp e^{-|\boldp| r} \,.
  \label{eq:potential_torus_6d}
\eeq
For general separations we perform the sum in eq.~\pref{eq:potential_torus_6d}
numerically, with results plotted as a function of $r$ in
fig.~\pref{fig:numeric_results}. As is clear from this figure, for large $r$
the interaction potential varies as $1/r$, as appropriate for 3 spatial
dimensions, while for small $r$ the sum over many KK modes combines to give a
result which varies as $1/r^3$, as is appropriate for forces in 5 spatial
dimensions. The various curves in this figure show how the results depend on
two of the moduli of the torus: the ratio $r_1/r_2$ of the lengths of its two
sides, as well as the `twist' angle $\theta$. For very asymmetric tori, with
one direction much longer than the other ({\it i.e.} small $r_1/r_2$ or small
$\theta$), the figure also shows that there can be an intermediate regime for
which there are effectively 4 spatial dimensions, and so for which the force
law varies as $1/r^2$.

\begin{figure}[h!]
  \centering
  \includegraphics[scale=1.1]{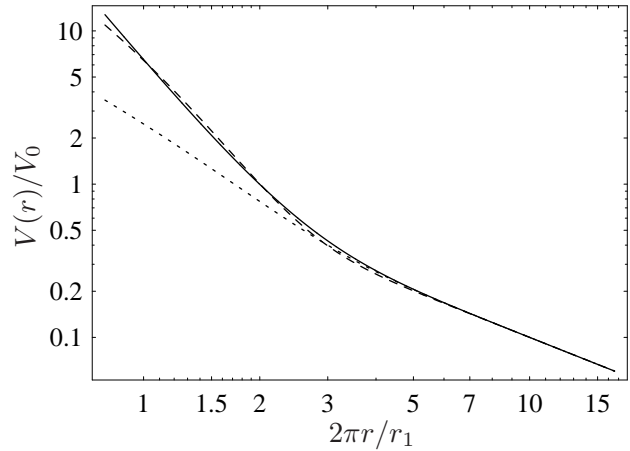}
  \caption{
  The exact potential for $\eta = 1$, $\theta = \pi/2$ (solid line) is fitted
  to an expression with a single exponential, of the form $(1 + a e^{-br})/r$.
  The best fit (dashed line), obtained from a least squares fit to the
  logarithm of $V/V_0$, is $(1 + 29.2481 \, e^{-1.68385 r})/r$, quite far from
  the "correct" expression $(1 + 4 e^{-r})/r$ from the first KK mode (dotted
  line). Here $r$ is written in units of $r_1 / (2\pi)$, and $V_0$ is defined
  in fig.~\pref{fig:numeric_results}.
  }
  \label{fig:singleexp}
\end{figure}

As the figures make clear, the crossover between the $1/r$ and
$1/r^2$ or $1/r^3$ power-law behaviour occurs over quite a short
range of $r$. This complicates the interpretation of experiments,
for which deviations to Newton's Law are often parameterized using
a single Yukawa-type exponential,
\be
  V(r) \sim \frac{1}{r} \left( 1 + a e^{-br} \right) \,,
\ee
with the parameters ($a$ and $b$) found by fitting to the data.
Naively, one might expect this form to provide a good description,
with the parameters found corresponding to the properties of the
first KK mode. In fact, as shown in fig.~\pref{fig:singleexp}, the
rapid crossover to a new power law implies the best fit is usually
quite different from what would be inferred from the first KK
mode, and also depends strongly on the interval of $r$'s used in
the fit. This emphasizes that some care is required when inferring
the form of the interaction potential directly from observational
data.

A final feature of these expressions which is worth mention is the
overall sign of the deviation from Newton's Law.
Non-supersymmetric 6D models involving only graviton exchange
generically predict that the strength of the interaction potential
{\it increases} relative to Newton's law as $r$ decreases, and
this prediction is robust because it follows only from the
attractive nature of graviton exchange.

The same need {\it not} be true in supersymmetric models because although
scalar and graviton exchange generate attractive forces between identical
particles, vector exchange leads to repulsion. In order to find a weaker force
than gravity as $r$ decreases it is therefore necessary to have the lowest
vector KK mode dominate over the relevant range of distances, which could in
principle be arranged by ensuring that the lightest KK mode correspond to a
bulk vector field. The above figures show that this does {\it not} occur when
these are compactified on a torus, however, because in this case all bulk
fields have identical KK spectra and so the vectors do not dominate the scalars
and gravitons. However this is an toroidal artefact, and a detection of an
initially negative residual could carry considerable information about the
shape of the extra dimensions.

\subsection{Extended Sources}

It is useful to generalize the result of
eq.~(\ref{eq:potential_torus_6d}), from point sources to extended
sources. This is a necessary complication because once the
interaction potential between point sources is no longer
proportional to $1/r$ it is not true that the potential due to a
spherical source distribution is the same as that of a point
charge.

In the present instance the interaction energy of two sources having mass
densities $\rho_1$ and $\rho_2$ whose centres are displaced by a separation
$\boldr$ is calculated by integrating eq.~\pref{eq:potential_torus_6d}, to give
\be
    V(r) = - \cG \sum_\boldp \int_{V_1} d^3 r_1 \int_{V_2} d^3 r_2 \,
    \rho_1(\boldr_1) \rho_2(\boldr_2)
    \frac{e^{-p|\boldr+\boldr_2-\boldr_1|}}{|\boldr+\boldr_2-\boldr_1|}
    \,.
\ee
Here
\beq
    \cG = G_N \left( 1 + \frac43 \langle X \rangle
    \right) - \alpha \omega_1 \omega_2 \,
\eeq
where the $\omega_i$ denote the charge-to-mass ratios for the two
source distribution, which is assumed to be constant. Similarly
$\langle X \rangle$ denotes the relevant average over the source
densities.

The density integrals can be calculated explicitly for simple
matter distributions, some of which we now perform for
illustrative purposes.

\subsubsection*{Spherical sources}

If we assume that the two sources are spheres with constant
density, $\rho_1$ and $\rho_2$, and radii, $R_1$ and $R_2$, so
$M_i = 4\pi \rho_i R_i^3/3$. We further assume $r \geq R_1 + R_2$
so that the two sources do not physically overlap. With these
choices the interaction potential energy becomes
\bea \label{eq:potential_twospheres}
    V(r) &=& -\cG \rho_1 \rho_2 \sum_\boldp
    \int_{r_1 \leq R_1} d^3 r_1 \int_{r_2 \leq R_2} d^3 r_2 \,
    \frac{e^{-p|\boldr+\boldr_2-\boldr_1|}}{|\boldr+\boldr_2-\boldr_1|}
    \nonumber\\
    &=& - \frac{\cG M_1 M_2}{r} \sum_\boldp
    B(|\boldp| R_1)\, B(|\boldp| R_2) \; e^{-|\boldp| r} \,,
\eea
where \cite{Jacques}
\bea
    B(x) = \frac{3 \left( x \cosh x - \sinh x \right)}{x^3}
    &\to& 1 + \frac{x^2}{10} + O(x^4) \qquad \qquad\qquad
    \hbox{if $x \ll 1$} \nonumber\\
    &\to& \frac{3 \, e^x}{2 \, x^2} \left( 1 - \frac{1}{x} \right) +
    O\left( e^{-x} \right)  \qquad \hbox{if $x \gg 1$}
    \, .
    \label{eq:B_sphere}
\eea

For each KK mode the effect of distributing the matter over a
sphere is to multiply each factor of mass, $M_i$, by the function
$B(|\boldp| R_i)$. Since $\lim_{x\to 0} B(x) = 1$ the contribution
of any KK mode whose wavelength, $\lambda = 1/|\boldp|$, is longer
than the sphere's radius is not modified from that of a point
source, and so in particular this is true for the KK zero modes.
Since these wavelengths are at their largest shorter than a micron
lengths, this means that for the spheres of practical interest the
macroscopic effects encoded in $B$ are important for all nonzero
KK modes. For these it is the opposite limit, $|\boldp|R \gg 1$,
which is relevant, in which case
\be \label{eq:twospheresRlarge}
    V(r) \approx - \frac{\cG M_1 M_2}{r} \left[ 1 + \frac{9}{4 R_1^2
    R_2^2} \sum_{\boldp \ne 0} \frac{1}{\boldp^4}
    \, e^{-|\boldp| (r-R_1-R_2)} \right]\,.
\ee
The sum over $\boldp$ is exponentially small unless the spheres are separated
by micron-sized distances of order the longest KK wavelength, $r - R_1 - R_2
\lesssim \lambda \sim \hbox{max} [1/|\boldp|]$, and for separations this large
it is $O(\lambda^4/R_1^2 R_2^2)$.

\begin{figure}[h!]
  \centering
  \includegraphics{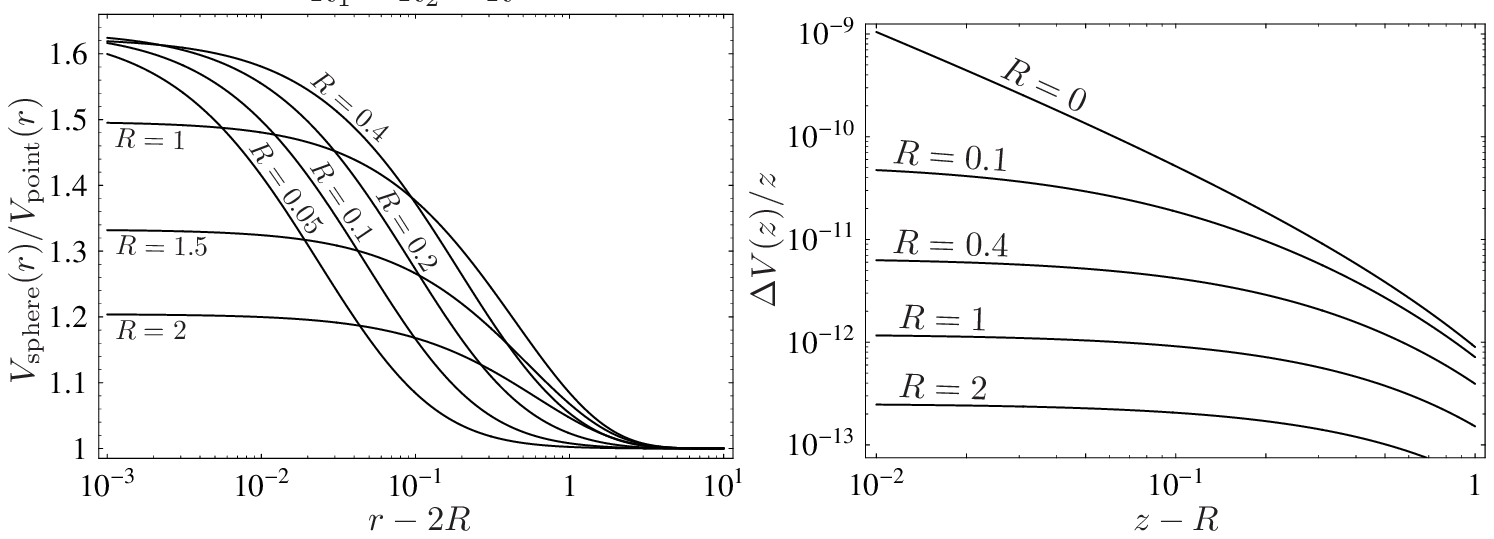}
  \caption{
  The figure to the left shows the ratio between the potential of spherical
  sources, eq.~\pref{eq:potential_twospheres}, and the corresponding point
  source potential. The radius is, when going from left to right and then down,
  $R = 0.05,\; 0.1,\; 0.2,\; 0.4,\; 1,\; 1.5\; \mathrm{and}\; 2$. The figure to
  the right shows the relative contribution of the KK sum
  in~\pref{eq:bulkplussphere}, for the potential between the Earth and a small
  sphere. The size of the torus is here chosen as $r_1 = 10^{-5}\; \mathrm{m}$.
  For both figures all lengths are written in units of $r_1 / (2\pi)$, and the
  moduli of the torus are chosen as $r_1 = r_2$ and $\theta = \pi/2$.
  }
  \label{fig:numeric_sphere}
\end{figure}

The sum in eq.~\pref{eq:potential_twospheres} is performed numerically and
compared to the point source potential in fig.~\pref{fig:numeric_sphere}, for
the simplest possible torus with $r_1 = r_2$ and $\theta = \pi/2$. The ratio
between the two potentials is always larger than $1$ since $B(x) \geq 1$. We
can find an upper bound to this ratio when the two spheres of equal radius
touch ($r = 2R$) by replacing the sums with integrals:
\be
  \frac{\sum_\boldp B^2(|\boldp| R) e^{-2|\boldp| R}}
    {\sum_\boldp e^{-2|\boldp| R}}
  \lesssim \int_0^\infty 4x B^2(x) e^{-2x} dx \approx 1.635\,532 \, .
\ee
This estimate is independent of the moduli of the torus.

\subsubsection*{Semi-Infinite Bulk Sources}

Of important practical interest is the case where a very small
spherical object interacts with a very large one (say, the Earth).
The potential energy in this case may be obtained from the above
result for spheres by taking the limit where the radius of one of
the spheres becomes very large. We are led in this way to the
interaction energy between a sphere of mass $M$ and radius $R$,
whose centre is displaced a distance $z$ from the edge of the
semi-infinite source.

Denoting the density and radius of the large sphere by $\rho$ and
$L$, we define its mass-per-unit-area as $\sigma = M/A = (4 \pi
\rho L^3/3)/(4 \pi L^2) = \rho L/3$. Dropping an overall additive
constant and neglecting sub-dominant powers of $z/L$ we find the
interaction energy to be
\be \label{eq:bulkplussphere}
    V(z) = M \left[ g  z - 2\pi \cG \rho
    \sum_{\boldp \ne 0}  B(|\boldp| R) \;
     \frac{1}{\boldp^2} \, e^{-|\boldp| z} \right] \,,
\ee
where $g = 4\pi \cG \sigma$. The first term in this result
describes the constant (Galilean) acceleration towards the large
source due to the exchange of massless KK modes, and since this
acceleration `sees' the entire source its strength depends on the
mass/area ratio $\sigma$.

By contrast, the second term represents an attraction of the smaller sphere
towards the nearest piece of the larger sphere due to the exchange of massive
KK states, which is only sensitive to the amount of matter within a range
$\lambda$ of the edge of the small sphere due to the finite range of these
interactions. Because of this finite range this part of the interaction is
controlled by $\rho$ rather than $\sigma$, and is exponentially small unless
the sphere's edge is within $\lambda$ ({\it i.e.} closer than a micron or so)
from the edge of the semi-infinite source (see fig.~\pref{fig:numeric_sphere}).
Because this attraction is only to nearby source material, it is also sensitive
to any local deviations of the large source from spherical symmetry.

The apparent divergence in eq.~\pref{eq:bulkplussphere} as $z \to
0$ (due to the divergence of the sum) is illusory so long as $R$
is nonzero. This is because the geometry requires $z \ge R$, with
the sphere touching the slab only when $z = R$. The result is
finite in this limit, as may also be seen explicitly from its
parent formula, eq.~\pref{eq:twospheresRlarge}.
Eq.~\pref{eq:bulkplussphere} would diverge in the limit $R \to 0$,
however, since in this case $B = 1$ and $z$ can approach zero.
This represents a divergence in the potential of a single
spherical which only arises at the source's edge, and as such it
may be absorbed into the coefficient of an effective interaction
which is localized on the boundary of the source.

In the limit of practical interest, $|\boldp|R \gg 1$, we may use
in eq.~\pref{eq:bulkplussphere} the asymptotic form for $B(x)$ to
obtain the simple formula
\be \label{eq:bulkplussphere1}
    V(z) = M \left[ g  z - \frac{3\pi \cG \rho}{R^2}
    \sum_{\boldp \ne 0}
     \frac{1}{\boldp^4} \, e^{-|\boldp| d} \right] \,,
\ee
where $d = z-R$ denotes the separation between the nearest edges
of the two sources.

\section{Conclusions}

The calculations in this paper allow the following conclusions to
be drawn concerning the implications of SLED models for tests of
gravity at short distances.

\begin{itemize}
\item SLED models predict deviations from the inverse-square law
at micron distance scales, giving instead a position-dependence which describes
a crossover from $1/r^2$ to $1/r^4$ over a fairly short range of distances.
These predictions are not well-described by perturbing Newton's Law with a
single Yukawa-type exponential potential. This situation is similar to what is
obtained in non-supersymmetric models with sub-millimeter extra dimensions,
with the important difference that within SLED the scale of the crossover is
related to the observed Dark Energy density and so cannot be adjusted to
smaller values if no deviations from Newton's Law are observed. It also differs
from the Yukawa-type deviations which are likely to be produced by recent
string models having a string scale in the TeV region with six large dimensions
\cite{Fernando}.
\item The precise shape of the deviations from Newton's Law in the
crossover region are likely to depend on the details of the
Kaluza-Klein spectrum, and so also on the precise shape of the
extra-dimensional geometry. This is already visible in the
calculations presented here, through the dependence of the forces
on the various shape moduli of the extra-dimensional torus.
Calculations are underway to determine more precisely how this
shape depends on these details for more general geometries than
the tori considered here.
\item SLED models also differ from their non-supersymmetric
counterparts in the number and kinds of fields which supersymmetry
dictates must populate the extra dimensions. In particular, since
bulk vector exchange mediates a repulsive force, it is in
principle possible to obtain an interaction energy which is weaker
than Newton's law at short distances if the lightest KK mode
should be for the vector fields rather than the tensors or
scalars. So far as we know this same possibility cannot arise
within the purely-gravitational extra-dimensional models
considered heretofore. This type of short-distance repulsive
interaction turns out never to dominate for the simple toroidal
compactifications considered here, but could be possible for more
complicated extra-dimensional geometries.
\item The presence of several bulk fields in SLED models can also
generically introduce a composition dependence to the observed
inter-particle force. Unfortunately, the detailed nature of this
composition-dependence is difficult to predict without knowing
more of the details of our extra-dimensional surroundings.
\end{itemize}

All of these conclusions underline the fact that the SLED proposal
is unique in that it provides a very precise framework within
which a fundamental connection is made between the observed Dark
Energy density and other kinds of observable physics. In it
deviations of gravity at sub-millimeter distances are also tied to
observable signals at the LHC and possibly to scalar-tensor
deviations from General Relativity over both astrophysical and
cosmological distance scales. We hope that experimenters take up
the challenge of searching for these phenomena.

\appendix

\section*{Acknowledgements}
This research is supported by a grant from N.S.E.R.C. (Canada) as
well as funds from the Killam Foundation, McMaster University and
the Perimeter Institute. The work of P.C. is supported by grant
no. NFR 153577/432 from the Research Council of Norway.


\begin{thebibliography}{99}


\bibitem{DEdiscovery}
S.~Perlmutter {\it et al.}, Ap.\ J.\ {\bf 483}, 565 (1997), [astro-ph/9712212];
%
A.G. Riess {\it et al.}, Ast.\ J.\ {\bf 116}, 1009 (1997), [astro-ph/9805201];
%
N. Bahcall, J.P. Ostriker, S. Perlmutter, P.J. Steinhardt, Science {\bf 284},
1481 (1999), [astro-ph/9906463].

\bibitem{ccreview}
S. Weinberg, Rev.\ Mod.\ Phys.\ {\bf 61}, 1 (1989).

\bibitem{anthropic}
S.~Weinberg,
  %``Anthropic Bound On The Cosmological Constant,''
  Phys.\ Rev.\ Lett.\ {\bf 59}, 2607 (1987).
  %%CITATION = PRLTA,59,2607;%%

\bibitem{anthropic2}
See, however, M.~L.~Graesser, S.~D.~H.~Hsu, A.~Jenkins and
M.~B.~Wise,
  %``Anthropic distribution for cosmological constant and primordial density
  %perturbations,''
  Phys.\ Lett.\ B {\bf 600}, 15 (2004),
  [hep-th/0407174];
  %%CITATION = HEP-TH 0407174;%%
%
 J.~Garriga and A.~Vilenkin,
  %``Anthropic prediction for Lambda and the Q catastrophe,''
  [hep-th/0508005].
  %%CITATION = HEP-TH 0508005;%%


\bibitem{landscape}
R.~Bousso and J.~Polchinski,
%``Quantization of four-form fluxes
%and dynamical neutralization of the  cosmological constant,''
JHEP {\bf 0006}, 006 (2000), [hep-th/0004134];
%
L.~Susskind,
%``The anthropic landscape of string theory,''
[hep-th/0302219].

\bibitem{splitsusy}
 L.~Susskind,
  %``Supersymmetry breaking in the anthropic landscape,''
  [hep-th/0405189];
  %%CITATION = HEP-TH 0405189;%%
%
 N.~Arkani-Hamed, S.~Dimopoulos, G.~F.~Giudice and A.~Romanino,
  %``Aspects of split supersymmetry,''
  Nucl.\ Phys.\ B {\bf 709}, 3 (2005),
  [hep-ph/0409232];
  %%CITATION = HEP-PH 0409232;%%
%
N.~Arkani-Hamed, S.~Dimopoulos and S.~Kachru,
  %``Predictive landscapes and new physics at a TeV,''
  [hep-th/0501082];
  %%CITATION = HEP-TH 0501082;%%
%
 M.~Dine, D.~O'Neil and Z.~Sun,
  %``Branches of the landscape,''
  JHEP {\bf 0507}, 014 (2005),
  [hep-th/0501214].
  %%CITATION = HEP-TH 0501214;%%



\bibitem{Towards}
Y. Aghababaie, C.P. Burgess, S. Parameswaran and F. Quevedo, Nucl.\ Phys.\ B
{\bf 680}, 389--414 (2004), [hep-th/0304256];
%%CITATION = HEP-TH 0304256;%%
%
Y. Aghabababie, C.P. Burgess, J.M. Cline, H. Firouzjahi, S. Parameswaran, F.
Quevedo, G. Tasinato and I. Zavala, JHEP {\bf 0309}, 037 (2003),
[hep-th/0308064];
%%CITATION = HEP-TH 0308064;%%
%
C.P. Burgess, F. Quevedo, G. Tasinato and I. Zavala,
[hep-th/0408109];
%%CITATION = HEP-TH 0408109;%%
%
%``Casimir Energies for 6D Supergravities Compactified on T2/ZN
%with Wilson Lines,''
D. Ghilencea, C.P. Burgess and F. Quevedo, [hep-th/0506164].
%%CITATION = HEP-TH 0506164;%%

\bibitem{Doug}
%
C.P. Burgess and D. Hoover, [hep-th/0504004];
%%CITATION = HEP-TH 0504004;%%
%
%``Ultraviolet Sensitivity in Higher Dimensions,''
D. Hoover and C.P. Burgess, [hep-th/0507293].
%%CITATION = HEP-TH 0507293;%%

\bibitem{SLEDreviews}
For reviews of the SLED proposal see:
%
C.P. Burgess, ``Supersymmetric Large Extra Dimensions and the Cosmological
Constant: An Update,'' Ann.\ Phys.\ {\bf 313}, 283--401 (2004),
[hep-th/0402200];
%%CITATION = HEP-TH 0402200;%%
%
%C.~P.~Burgess,
and ``Towards a natural theory of dark energy: Supersymmetric
large extra dimensions,'' in the proceedings of the Texas A\&M
Workshop on String Cosmology, [hep-th/0411140].
%%CITATION = HEP-TH 0411140;%%

\bibitem{ADD}
N. Arkani-Hamed, S. Dimopoulos and G. Dvali, Phys.\ Lett.\ B {\bf 429}, 263
(1998), [hep-ph/9803315]; Phys.\ Rev.\ D {\bf 59}, 086004 (1999),
[hep-ph/9807344];
%
I.~Antoniadis, N.~Arkani-Hamed, S.~Dimopoulos and G.~R.~Dvali,
%``New dimensions at a millimeter to a Fermi and superstrings at a TeV,''
Phys.\ Lett.\ B {\bf 436}, 257 (1998), [hep-ph/9804398].

\bibitem{NLTests}
C. D. Hoyle, U. Schmidt, B. R. Heckel, E. G. Adelberger, J. H. Gundlach, D. J.
Kapner, H. E. Swanson, Phys.\ Rev.\ Lett.\ {\bf 86}, 1418--1421 (2001),
[hep-ph/0011014];
%
S.K. Lamoreaux, Phys.\ Rev.\ Lett.\ {\bf 78}, 5 (1997);
%
M. Bordag, B. Geyer, G.L. Klimchitskaya and V.M. Mostepanenko, Phys.\ Rev.\ D
{\bf 58}, 075003 (1998);
%
J. Chiaverini, S.J. Smullin, A.A. Geraci, D.M. Weld and A. Kapitulnik, Phys.\
Rev.\ Lett.\ {\bf 90}, 151101 (2003), [hep-ph/0209325];
%
J.C. Long, H.W. Chan, A.B. Churnside, E.A. Gulbis, M.C.M. Varney and J.C.
Price, Nature {\bf 421}, 922 (2003);
%
C.D. Hoyle, D.J. Kapner, B.R. Heckel, E.G. Adelberger, J.H. Gundlach, U.
Schmidt and H.E. Swanson, Phys.\ Rev.\ D {\bf 70}, 042004 (2004),
[hep-ph/0405262].

\bibitem{NLTestsRev}
For a recent review with references, see
%
E.G. Adelberger, B.R. Heckel and A.E. Nelson, Ann.\ Rev.\ Nucl.\ Part.\ Sci.\
{\bf 53}, 77--121 (2003), [hep-ph/0307284].

\bibitem{Precursors}
For some early precursors to these ideas see:
%
V.A. Rubakov and M.E. Shaposhnikov, Phys.\ Lett.\ B {\bf 125}, 139 (1983);
%
C. Wetterich, Nucl.\ Phys.\ B {\bf 255}, 480 (1985).

\bibitem{SLEDrelated}
S.~M.~Carroll and M.~M.~Guica,
%``Sidestepping the cosmological constant with football-shaped extra
%dimensions,''
[hep-th/0302067];
%
%%CITATION = HEP-TH 0302067;%%
%\bibitem{navarro}
I.~Navarro,
%``Codimension two compactifications and the cosmological constant  problem,''
JCAP {\bf 0309}, 004 (2003), [hep-th/0302129];
%%CITATION = HEP-TH 0302129;%%
%
I.~Navarro,
%``Spheres, deficit angles and the cosmological constant,''
Class.\ Quant.\ Grav.\ {\bf 20}, 3603 (2003), [hep-th/0305014];
%%CITATION = HEP-TH 0305014;%%
%
G.~W.~Gibbons, R.~Guven and C.~N.~Pope,
%``3-branes and uniqueness of the Salam-Sezgin vacuum,''
Phys.\ Lett.\ B {\bf 595}, 498 (2004), [hep-th/0307238];
%%CITATION = HEP-TH 0307238;%%
%
H.~P.~Nilles, A.~Papazoglou and G.~Tasinato,
%``Selftuning and its footprints,''
Nucl.\ Phys.\ B {\bf 677}, 405 (2004), [hep-th/0309042];
%%CITATION = HEP-TH 0309042;%%
%
I.~Navarro and J.~Santiago,
%``Higher codimension braneworlds from intersecting branes,''
JHEP {\bf 0404}, 062 (2004), [hep-th/0402204];
%%CITATION = HEP-TH 0402204;%%
%
A.~L.~Maroto,
%``Brane oscillations and the cosmic coincidence problem,''
Phys.\ Rev.\ D {\bf 69}, 101304 (2004), [hep-ph/0402278];
%%CITATION = HEP-PH 0402278;%%
%
S.~Randjbar-Daemi and E.~Sezgin,
%``Scalar potential and dyonic strings in 6d gauged supergravity,''
Nucl.\ Phys.\ B {\bf 692}, 346 (2004), [hep-th/0402217];
%%CITATION = HEP-TH 0402217;%%
%
M.~L.~Graesser, J.~E.~Kile and P.~Wang,
%``Gravitational perturbations of a six dimensional self-tuning model,''
Phys.\ Rev.\ D {\bf 70}, 024008 (2004), [hep-th/0403074];
%%CITATION = HEP-TH 0403074;%%
%
C.~Kim, Y.~Kim and O.~K.~Kwon,
%``Tubular D-branes in Salam-Sezgin model,''
JHEP {\bf 0405}, 020 (2004), [hep-th/0404163];
%%CITATION = HEP-TH 0404163;%%
%
I.~Navarro and J.~Santiago,
%``Flux compactifications: Stability and implications for cosmology,''
JCAP {\bf 0409}, 005 (2004), [hep-th/0405173];
%%CITATION = HEP-TH 0405173;%%
%
A.~Kehagias,
%``A conical tear drop as a vacuum-energy drain for the solution of the
%cosmological constant problem,''
Phys.\ Lett.\ B {\bf 600}, 133 (2004), [hep-th/0406025];
%%CITATION = HEP-TH 0406025;%%
%
J.~Vinet and J.~M.~Cline,
%``Can codimension-two branes solve the cosmological constant problem?,''
[hep-th/0406141];
%%CITATION = HEP-TH 0406141;%%
%
and [hep-th/0501098];
%%CITATION = HEP-TH 0501098;%%
%
J. Garriga and M. Porrati, [hep-th/0406158];
%
T.~Biswas and P.~Jaikumar,
%``Cosmology from moduli dynamics,''
JHEP {\bf 0408}, 053 (2004), [hep-th/0407063];
%%CITATION = HEP-TH 0407063;%%
%
S.~Randjbar-Daemi and V.~Rubakov,
%``4d-flat compactifications with brane vorticities,''
[hep-th/0407176];
%%CITATION = HEP-TH 0407176;%%
%
H.~M.~Lee and A.~Papazoglou,
%``Brane solutions of a spherical sigma model in six dimensions,''
[hep-th/0407208];
%
%%CITATION = HEP-TH 0407208;%%
V.P. Nair and S.~Randjbar-Daemi, [hep-th/0408063];
%
R.~Erdem,
%``A symmetry for vanishing cosmological constant in an extra dimensional toy
%model,''
[hep-th/0410063];
%%CITATION = HEP-TH 0410063;%%
%
S.~Nobbenhuis,
%``Categorizing different approaches to the cosmological constant problem,''
[gr-qc/0411093];
%%CITATION = GR-QC 0411093;%%
%
P.~Brax, C.~van de Bruck and A.~C.~Davis,
%``Cosmic acceleration in massive half-maximal supergravity,''
[hep-th/0411208];
%%CITATION = HEP-TH 0411208;%%
%
I.~Navarro and J.~Santiago,
%``Gravity on codimension 2 brane worlds,''
[hep-th/0411250];
%%CITATION = HEP-TH 0411250;%%
%
%\cite{Kainulainen:2004vk}
  K.~Kainulainen and D.~Sunhede,
  %``Dark energy from large extra dimensions,''
  [astro-ph/0412609];
  %%CITATION = ASTRO-PH 0412609;%%
%
M.~Redi,
%``Footballs, conical singularities and the Liouville equation,''
Phys.\ Rev.\ D {\bf 71}, 044006 (2005), [hep-th/0412189];
%%CITATION = HEP-TH 0412189;%%
%
G.~Kofinas,
%``Conservation equation on braneworlds in six dimensions,''
[hep-th/0412299];
%%CITATION = HEP-TH 0412299;%%
%
J.~M.~Schwindt and C.~Wetterich,
%``Dark energy cosmologies for codimension-two branes,''
[hep-th/0501049];
%%CITATION = HEP-TH 0501049;%%
%
E.~Papantonopoulos and A.~Papazoglou,
%``Brane-bulk matter relation for a purely conical codimension-2 brane
%world,''
[hep-th/0501112];
%%CITATION = HEP-TH 0501112;%%
%
%\cite{Charmousis:2005ey}
C.~Charmousis and R.~Zegers,
%``Matching conditions for a brane of arbitrary codimension,''
[hep-th/0502170];
%%CITATION = HEP-TH 0502170;%%
%
 H.~M.~Lee and C.~Ludeling,
%``The general warped solution with conical branes in six-dimensional
%supergravity,''
[hep-th/0510026].
%%CITATION = HEP-TH 0510026;%%

\bibitem{LEDPheno}
G.F. Giudice, R. Rattazzi and J.D. Wells, Nucl.\ Phys.\ B {\bf 544}, 3--38
(1999), [hep-ph/9811291];
%
E.A. Mirabelli, M. Perelstein and M.E. Peskin, Phys.\ Rev.\ Lett.\ {\bf 82},
2236--2239 (1999), [hep-ph/9811337];
%
J.L. Hewett, Phys.\ Rev.\ Lett.\ {\bf 82}, 4765--4768 (1999), [hep-ph/9811356];
%
T Han, J.D. Lykken and R.-J. Zhang, Phys.\ Rev.\ D {\bf 59}, 105006 (1999),
[hep-ph/9811350];
%
I. Antoniadis, K. Benakli and M. Quiros, Phys.\ Lett.\ B {\bf 460}, 176--183
(1999), [hep-ph/9905311];
%
S. Cullen, M. Perelstein and M.E. Peskin, Phys.\ Rev.\ D {\bf 62}, 055012
(2000), [hep-ph/0001166].

\bibitem{preSLEDPheno}
D. Atwood, C.P. Burgess, E. Filotas, F. Leblond, D. London and I. Maksymyk,
Phys.\ Rev.\ D {\bf 63}, 025007 (2001), [hep-ph/0007178];
%%CITATION = HEP-PH 0007178;%%
%
J.~L.~Hewett and D.~Sadri,
%``Supersymmetric extra dimensions: Gravitino effects in selectron pair
%production,''
Phys.\ Rev.\ D {\bf 69}, 015001 (2004).
%%CITATION = HEP-PH 0204063;%%


\bibitem{MSLED}
%
C.P. Burgess, J. Matias and F. Quevedo, [hep-ph/0404135].

\bibitem{SLEDPheno}
%
G. Azuelos, P.H. Beauchemin and C.P. Burgess, [hep-ph/0401125];
%%CITATION = HEP-PH 0401125;%%
%
P.H. Beauchemin, G. Azuelos and C.P. Burgess, J.\ Phys.\ G {\bf 30}, N17
(2004), [hep-ph/0407196];
%%CITATION = HEP-PH/0407196;%%
%
J. Matias and C.P. Burgess,
%``MSLED, Neutrino Oscillations and the Cosmological Constant,''
JHEP {\bf 0509}, 052 (2005), [hep-ph/0508156].
%%CITATION = HEP-PH/0508156;%%


\bibitem{ABRS}
A. Albrecht, C.P. Burgess, F. Ravndal and C. Skordis, Phys.\ Rev.\ D {\bf 65},
123506 (2002), [astro-ph/0107573].
%%CITATION = ASTRO-PH 0107573;%%

\bibitem{SLEDCosmo}
A. Albrecht and C. Skordis, Phys.\ Rev.\ Lett.\ {\bf 84}, 2076 (2000),
[astro-ph/9908085];
%
M. Peloso and E. Poppitz, Phys.\ Rev.\ D {\bf 68}, 125009 (2003),
[hep-ph/0307379];
%
K.~Kainulainen and D.~Sunhede,
  %``Dark energy from large extra dimensions,''
  [astro-ph/0412609].
  %%CITATION = ASTRO-PH 0412609;%%


\bibitem{Jacques}
C.P. Burgess and J. Cloutier, Phys.\ Rev.\ D {\bf 38}, 2944--2950 (1988).

\bibitem{WbgG&C}
S.~Weinberg, {\sl Gravitation and Cosmology}, Wiley, New York,
1972.

\bibitem{MTW}
C.W.~Misner, K.S.~Thorne and J.A.~Wheeler, {\sl Gravitation}, W.H. Freeman and
Co., 1973.

\bibitem{FVTerms}
 B.~Holdom,
  %``Two U(1)'S And Epsilon Charge Shifts,''
  Phys.\ Lett.\ B {\bf 166}, 196 (1986);
  %%CITATION = PHLTA,B166,196;%%
%
D. Atwood, C.P. Burgess, E. Filotas, F. Leblond, D. London and I. Maksymyk,
Phys.\ Rev.\ D {\bf 63}, 025007 (2001), [hep-ph/0007178];
%%CITATION = HEP-PH 0007178;%%
%
 S.~A.~Abel and B.~W.~Schofield,
  %``Brane-antibrane kinetic mixing, millicharged particles and SUSY
  %breaking,''
  Nucl.\ Phys.\ B {\bf 685}, 150 (2004),
  [hep-th/0311051];
  %%CITATION = HEP-TH 0311051;%%
%
  S.~Abel and J.~Santiago,
  %``Constraining the string scale: from Planck to Weak and back again,''
  J.\ Phys.\ G {\bf 30}, R83 (2004),
  [hep-ph/0404237].
  %%CITATION = HEP-PH 0404237;%%

\bibitem{NS}
H. Nishino and E. Sezgin, Phys.\ Lett.\ B {\bf 144}, 187 (1984);
%``The Complete N=2, D = 6 Supergravity With Matter And Yang-Mills
%Couplings,''
Nucl.\ Phys.\ B {\bf 278}, 353 (1986);
%%CITATION = NUPHA,B278,353;%%
%
S. Randjbar-Daemi, A. Salam, E. Sezgin and J. Strathdee, Phys.\ Lett.\ B {\bf
151}, 351 (1985); A.~Salam and E.~Sezgin,
%``Chiral Compactification On Minkowski X S**2 Of N=2 Einstein-Maxwell
%Supergravity In Six-Dimensions,''
Phys.\ Lett.\ B {\bf 147}, 47 (1984);
%%CITATION = PHLTA,B147,47;%%
%
L.J. Romans, Nucl.\ Phys.\ B {\bf 269}, 691--711 (1986).

\bibitem{Fernando}
  V.~Balasubramanian, P.~Berglund, J.~P.~Conlon and F.~Quevedo,
  %``Systematics of moduli stabilisation in Calabi-Yau flux compactifications,''
  JHEP {\bf 0503}, 007 (2005),
  [hep-th/0502058];
  %%CITATION = HEP-TH 0502058;%%
%
J.~P.~Conlon, F.~Quevedo and K.~Suruliz,
  %``Large-volume flux compactifications: Moduli spectrum and D3/D7 soft
  %supersymmetry breaking,''
  JHEP {\bf 0508}, 007 (2005),
  [hep-th/0505076].
  %%CITATION = HEP-TH 0505076;%%


\end{thebibliography}
\end{document}